\documentclass[3p,times]{elsarticle}
\usepackage{pgf-pie}
\usepackage{array}
\usepackage{ecrc}
\usepackage{url}
\usepackage{multirow}
\usepackage{tabularx}
\usetikzlibrary{patterns}
\usetikzlibrary{decorations.text}

\volume{00}

\firstpage{1}

\journalname{Blockchain:Research and Applications}

\runauth{Kivilo, S.; Norta, A.; Hattingh, M.; Avanzo, S.; Pennella, L}

\jid{Res Appl}

\jnltitlelogo{Blockchain: R and A}

\CopyrightLine{2026}{Published by Elsevier Ltd.}

\usepackage{amssymb}
\usepackage[T1]{fontenc}

\usepackage[figuresright]{rotating}

\begin{document}

\begin{frontmatter}



\dochead{}

\title{Designing a Token Economy: Incentives, Governance, and Tokenomics}

\author[1]{Samela Kivilo}
\author[2,4]{Alex Norta\corref{cor1}\fnref{fn1}}
\author[3]{Marie Hattingh}
\author[5]{Sowelu Avanzo}
\author[6]{Luca Pennella}

\cortext[cor1]{Corresponding author}
\fntext[fn1]{ORCID: 0000-0003-0593-8244}

\address[1]{Tallinn University of Technology, Estonia; samela.kivilo@gmail.com}
\address[2]{Tallinn University, Estonia; alex.norta.phd@ieee.org}
\address[3]{University of Pretoria, South Africa; marie.hattingh@up.ac.za}
\address[4]{Dymaxion OÜ, Tallinn, Estonia}
\address[5]{University of Torino, Italy; soweluelios.avanzo@unito.it}
\address[6]{University of Trieste, Italy; luca.pennella@phd.units.it}

\begin{abstract}
In recent years, tokenomic systems, decentralized systems that use cryptographic tokens to represent value and rights, have evolved considerably. Growing complexity in incentive structures has expanded the applicability of blockchain beyond purely transactional use. Existing research predominantly examines token economies within specific use cases, proposes conceptual frameworks, or studies isolated aspects such as governance, incentive design, and tokenomics. However, the literature offers limited empirically grounded, end-to-end guidance that integrates these dimensions into a coherent, step-by-step design approach informed by concrete token-economy development efforts. To address this gap, this paper presents the Token Economy Design Method (TEDM), a design-science artifact that synthesizes stepwise design propositions for token-economy design across incentives, governance, and tokenomics. TEDM is derived through an iterative qualitative synthesis of prior contributions and refined through a co-designed case. The artifact is formatively evaluated via the Currynomics case study and additional expert interviews. Currynomics is an ecosystem that maintains the Redcurry stablecoin, using real estate as the underlying asset. TEDM is positioned as reusable design guidance that facilitates the analysis of foundational requirements of tokenized ecosystems. The specificity of the proposed approach lies in the focus on the socio-technical context of the system and early stages of its design.
\end{abstract}


\end{frontmatter}



\section{Introduction}\label{introduction}
Blockchain technology has become increasingly popular in recent years due to its ability to provide decentralization, security, and transparency. Still, even with significant advances in blockchain technology over time, its true potential has not yet been fully realized. Although simple cryptocurrency transfers are straightforward to execute and comprehend, it remains challenging to link such transactions to real-world objects or activities. The latter makes it difficult to expand the range of beneficial use cases for blockchains~\cite{Santos2021}. As a solution, cryptographic tokens have emerged that can be used to represent various assets or services within a network~\cite{Oliviera2018}.

Before blockchain technology, tokens were commonly associated with vouchers that indicate tangible values such as casino chips or beverages at festivals~\cite{Schubert2021}. In the blockchain context, tokens act as a claim on both physical and immaterial assets, e.g., gold ingots, real estate, and the accompanying rental contracts. Establishing a connection between real-life technology and blockchain paves the way for the development of advanced and innovative applications. Thus, tokens play an important role in offering competition to hyperscaler web2 platforms that have reached a near-monopoly status during the last decade~\cite{liu2022}. These dominant platforms, for example, Facebook, Instagram, Airbnb and Google, own user data in a centralized way, allowing the former to use them to increase proprietary profits~\cite{Aistov2020}. Such centralization threatens the privacy of user data and increases the general vulnerability of the ecosystem against hackers~\cite{Khamisa2021}. On the other hand, tokens enable translating various functionalities of such platforms, from social networks to search engines and file storage, into a decentralized form, where user data cannot be maliciously converted into profits for intermediary platforms.

In terms of blockchain startups, tokens are an innovative channel of financing and can function as an ``internal currency'' of ecosystems~\cite{Pietrewicz2018}. Additionally, tokens enable novel organizational forms often referred to as decentralized autonomous organizations (DAOs). In practice, DAOs differ substantially in their degree of automation and decentralization: some rely primarily on on-chain smart contracts and token-based governance, while others combine on-chain execution with off-chain coordination, social processes, and informal norms. In this context, tokens frequently provide the basis for membership and governance mechanisms, and token-based voting is adopted in a large share of governance protocols~\cite{kurniawanvoting, baninemeh2023decision, ziegler2022taxonomy}. This interlinks decision making in DAOs with key functions such as conditional value storage and transfer, which are essential for DAO operations~\cite{rikken2023ins}, and underscores the importance of designing sound token economies for DAOs.

The literature review in Section~\ref{sec:literaturereview} suggests three recurring design dimensions in token economies: incentives, governance, and tokenomics. Incentives concern rewards and penalties that shape participant behavior toward system goals. Governance refers to how decision-making rights are allocated and exercised. Tokenomics defines the rules for token issuance, distribution, allocation, and burning~\cite{Benedetti2023}. In this work, we focus on these dimensions as a pragmatic structuring lens for design, while acknowledging that additional technical choices may condition feasible solutions.

Several studies highlighted the relevance of token economy design for blockchain ecosystems and their governance~\cite{wang2023blockchain, Han2022}.
Tokens in this context can align the interests of various stakeholder groups in a blockchain ecosystem, since the value of the token is related to the inherent value added to the ecosystem itself~\cite{Khamisa2021}. Yet, the wide range of token functionalities raises the need for careful tokenomic design. Developers often do not fully understand the incentive structures they create for users of blockchain systems, as well as how incentivization can backfire in the emergence of unexpected market events~\cite{Biancotti2022}. For example, ill-reasoned use of combined token functionalities, asset and payment tokens, can hinder the growth of token ecosystems~\cite{Drasch2020}. Furthermore, excessive token liquidity decreases the market capitalization of a token economy with respect to future profits, setting a financing limit~\cite{Chod2021}. Sockin \& Xiong~\cite{Sockin2020} warn against a token price collapse, a situation where there is no equilibrium price that would match the token supply with the demand, commonly caused by speculation.

One of the largest empirical examples of poor tokenomic design is the collapse of Luna and TerraUSD tokens in 2022, causing thousands of stakeholders to lose their investments~\cite{Faux2022}. Blockchain collapses were caused by the excessive accumulation of risk within the cryptocurrency ecosystem~\cite{Biancotti2022}. That is, inadequate risk management often coincided with poor product design, so tokens were created in a way that left them vulnerable to significant losses when unexpected events occurred. In 2022, many digital asset initiatives and token releases continued to cause damage to the people who used them and those who bought the tokens~\cite{Kaal2022}. These individuals often ended up with nothing more than worthless tokens after an initial period of success.

Prior work proposes several conceptual frameworks for token economies (e.g.,~\cite{Lage2022,Barrera2020,Benedetti2023,Khamisa2021}) and a growing body of domain-specific design discussions (e.g., industrial, automotive, Web3 gaming)~\cite{Wang2020,Hofmann2021,Zhang2019,Direr2022}. However, end-to-end guidance that (i) integrates incentives, governance, and tokenomics in a single coherent structure, (ii) provides stepwise design propositions with explicit decision points and trade-offs, and (iii) is grounded in empirical evidence from concrete token-economy development efforts remains limited. Moreover, several studies address one of the three dimensions in isolation, such as governance~\cite{Bena2022} or tokenomics in ICO settings~\cite{Gan2022}, while incentive-mechanism modeling is often treated separately~\cite{han2022can}.

In the domain of DAOs, research emphasizes the importance of token economy design for system utility, with references to architecture models that incorporate incentive mechanisms~\cite{qin2022web3, wang2019decentralized}. Furthermore, the need for a comprehensive software engineering method for token economy design in DAOs is highlighted in~\cite{avanzo2023modelling}, along with the support tool to validate DAO incentive mechanisms, which are proposed in~\cite{liang2023cadcad}. However, a complete software engineering method for this scope is not yet developed.

Taking these challenges into account, the objective of this study is to propose a Token Economy Design Method (TEDM) as a design-science artifact consisting of structured design propositions grounded in an empirical token-economy development case. We follow a Design Science Research (DSR) approach in which iterative design cycles address the following research questions~\cite{designResearch04}:
\begin{itemize}
\item \textbf{Main Research Question:} \textit{How can a token economy be designed in a structured way by explicitly integrating incentives, governance, and tokenomics?}
\item \textbf{Subresearch Question 1 (RQ1):} \textit{What design considerations and decision points support incentive-mechanism design in token economies?}
\item \textbf{Subresearch Question 2 (RQ2):} \textit{What design considerations and decision points support governance design in token economies?}
\item \textbf{Subresearch Question 3 (RQ3):} \textit{What design considerations and decision points support tokenomics design in token economies?}
\end{itemize}

To this end, Section~\ref{sec:methdology} describes the proposed DSR approach. The existing literature is examined in Section~\ref{sec:literaturereview} to identify the theoretical frameworks overlapping for token economy design and to introduce the three pillars of the proposed design artifact. In Section~\ref{sec:usecase} the Currynomics token system use case is introduced that helps the search process for effective design principles. The Token Economy Design Method (TEDM) is introduced in Section~\ref{sec:tokeneconomydesignmodel}. Each of the components of the design artifact is expanded upon in Sections~\ref{sec:TEDMincentives},~\ref{sec:TEDMgovernance},~\ref{sec:TEDMtokenomics}, respectively, to present the results of each sub-research question. Section~\ref{sec:evaluation} presents a discussion of how the artifact was evaluated, first, through a case study that evaluates the real-life practicality of the proposed design guidelines when used by the Currynomics ecosystem and, second, additional expert interviews. Finally, Section~\ref{conclusion} concludes the study with recommendations for future research. 

\section{Preliminaries} \label{sec:preliminaries}
This section provides the background information necessary to understand the main contributions of the article. Firstly, in Section \ref{sec:methdology}, the methodology adopted for the study and the data collection and analysis methods are described. Subsequently, Section \ref{sec:literaturereview} presents the related work, and Section \ref{sec:usecase} describes the use case.
\subsection{Methodology and Data}\label{sec:methdology}
The methodology chosen for this study is Design Science Research (DSR), as it is well suited for sociotechnical systems such as token economies~\cite{designResearch04}. 
The purpose of the study is to develop and evaluate a Token Economy Design Method (TEDM) that supports the early-stage structuring of token-economy design decisions across incentives, governance, and tokenomics. The DSR artifact consists of a practical guide that represents the core structure of the method and is informed by concepts derived from extant literature (Section~\ref{sec:literaturereview}) and refined through an empirical design case.

\begin{figure}
    \centering
    \includegraphics[scale = 0.35]{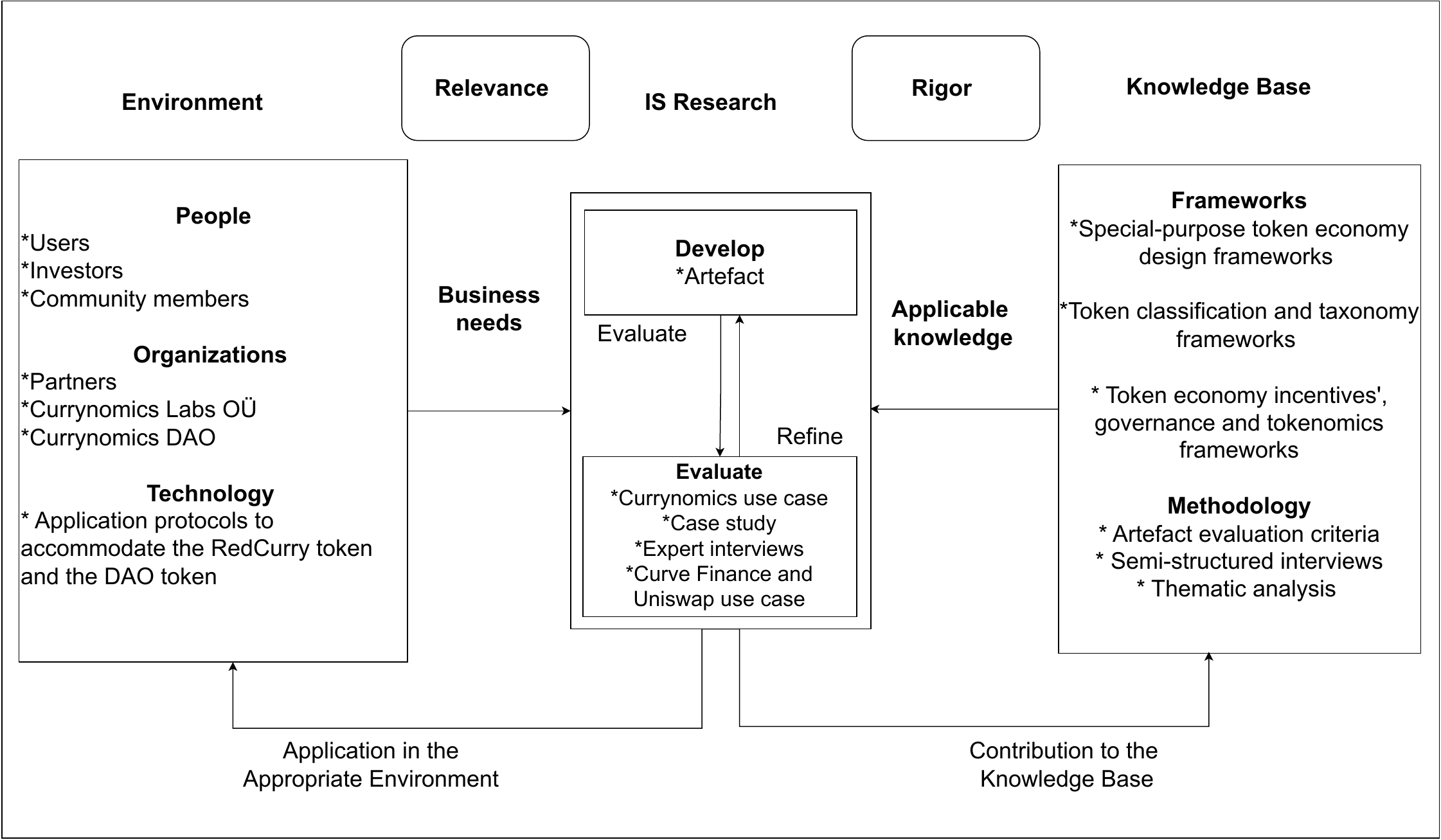}
    \caption{Design Science Research (DSR) framework adopted in this study~\cite{designResearch04}.}
    \label{fig:DSR}
\end{figure}

In line with DSR, the knowledge base of the study includes (i) foundational theories and frameworks relevant to token economies (Section~\ref{sec:theoreticalframeworks}) and (ii) established research methodologies and design-oriented approaches used for adjacent problems in the literature (Sections~\ref{sec:governance}, \ref{sec:incentives}, and \ref{sec:tokenomics}). Based on this knowledge base, we identify three key design pillars that organize the method: incentive structures, governance, and tokenomics. Importantly, TEDM is not intended as a predictive model of economic performance. Rather, it operationalizes empirically grounded design propositions that make decision points, trade-offs, and risks explicit, thereby supporting analytical reasoning and communication during token-economy development.
Figure~\ref{fig:DSR} illustrates the DSR research approach followed in this study. 
DSR aims to ensure the relevance of developed artifacts by targeting concrete business needs.
The environmental component of the DSR study captures key needs, shortcomings, and objectives of the focal organization and identifies resource constraints. Consistent with DSR, the contribution of this work lies in articulating a reusable design artifact grounded in empirical evidence. The evaluation is formative and supports analytical transferability (i.e., applicability-by-reasoning across contexts), rather than statistical generalization or causal proof of superior economic outcomes.
Our case study is focused on the Currynomics ecosystem, whose environment includes stakeholders (users, investors, and community members), organizations (partners, Currynomics Labs OÜ, and Currynomics DAO), and the technology utilized in the form of application protocols to accommodate the Redcurry token and the DAO token. Each of these elements is described in detail in Section~\ref{sec:usecase}.

\subsubsection{Artifact derivation and traceability}\label{sec:derivation}
The TEDM was derived through an iterative and qualitative process consistent with Design Science Research (DSR) for sociotechnical systems~\cite{designResearch04}. First, we conducted a scoped synthesis of extant research on token economy design, focusing on recurring constructs, decision points, and trade-offs reported across incentives, governance, and tokenomics (Section~\ref{sec:literaturereview}). Second, we performed concept extraction and clustering to consolidate overlapping constructs into a coherent set of design propositions. This synthesis resulted in three macro-dimensions that repeatedly co-occur in prior work and in practice: incentive structures, governance, and tokenomics. Third, the resulting propositions were iteratively refined through design cycles with the Currynomics case, where each step was checked for (i) goal clarity, (ii) operational interpretability, and (iii) traceable grounding in either literature, case evidence, or both. 
Accordingly, the TEDM should be interpreted as empirically grounded, reusable design guidance with explicit boundary conditions, rather than a statistically generalizable predictive model of token-economy outcomes.

\subsubsection{Data Collection and Analysis}\label{datacollection}

In DSR artifact evaluation, expert interviews can be an effective method for eliciting informed judgments on the artifact's utility and interpretability~\cite{Wieringa2014}. Semi-structured interviews are a valuable means of collecting data by getting insights from experts about their practices, beliefs, experiences, and points of view~\cite{Harrell2009}. Each TEDM component (Sections~\ref{sec:TEDMincentives}, \ref{sec:TEDMgovernance}, and \ref{sec:TEDMtokenomics}) underwent four iterative cycles of development, evaluation, and refinement. In the first cycle, concepts derived from extant literature were consolidated into initial design propositions and arranged into a stepwise method. The artifact was then evaluated through a use-case \textit{demonstration}, i.e., a simplified evaluation that shows how an artifact can be used to address a specific problem instance~\cite{Prat2015}. This step involved applying the prescribed TEDM steps to the Currynomics case. A purposive sample of two Currynomics representatives were first introduced to the visual method steps and subsequently instantiated the goal model shown in Figure~\ref{fig:tokeneconomydesignmodel} based on project requirements. The artifact utility was then evaluated through semi-structured interviews with the two Currynomics representatives (IN1, IN2). Additional feedback was collected from a distinct purposive sample of three industry experts external to the case (IN3, IN4, IN5).
A final evaluation step consisted of applying the TEDM categories to compare two existing token economies as a comparative demonstration, aiming to assess whether the method's constructs and decision points remain interpretable beyond the focal case. This step does not test economic performance outcomes, but supports analytical transferability of the method's structuring logic across contexts. 

The objective of the interviews is to evaluate the artifact's \textit{completeness}, \textit{simplicity}, \textit{understandability}, \textit{operational feasibility}, and perceived \textit{accuracy}, evaluation criteria commonly adopted in IS/DSR artifact evaluation~\cite{Prat2015}. The selection of these criteria is inspired by van Pelt et al.~\cite{vanPelt2021}, who developed and evaluated a blockchain governance framework with a narrower scope than the present study. Interviews were transcribed using the Transkriptor software\footnote{https://app.transkriptor.com/uploader}. The collected data were analyzed via thematic analysis following the six steps outlined by Braun and Clarke~\cite{Braun2014}, using the coding functionality of MAXQDA\footnote{https://www.maxqda.com/}. At the beginning of each interview, participants were informed about the purpose of the study and the interview structure, and consent to record the interview was requested. In the first part, participants were introduced to TEDM and invited to comment on any design topic they considered salient. In the second part, the following questions were asked:
\begin{enumerate}
    \item How do you perceive the completeness of the method?
    \item How would you comment on the simplicity of the method?
    \item How do you perceive the understandability of the method?
    \item What are your thoughts on the operational feasibility of the method?
    \item What are your thoughts on the perceived accuracy of the method?
\end{enumerate}
In Table \ref{table:goal_model_notation}, we present the notation adopted in the use case examples to illustrate the method. The notation, which is part of Agent-Oriented Modeling (AOM), was chosen due to its simplicity and explainability, as well as its suitability for sociotechnical systems such as token economies \cite{sterling2009art}. In AOM, goal models are used to represent and analyze system requirements. The goals that agents pursue correspond to the functional requirements of the system, while the quality goals correspond to the quality attributes associated with the given functions. Goals are hierarchically decomposed into sub-goals. 
The following section presents a review of the literature that contextualizes the foundation of the method.
\begin{table}[ht]
\centering
\begin{tabular}{|c|p{5.9cm}|c|}
\hline
\textbf{Elements} & \textbf{Description} & \textbf{Notation} \\
\hline
\textbf{Goal} & A particular situation or state of things in the world that the stakeholders wish to attain. & 
\makebox[2cm][c]{\vspace{5pt}\includegraphics[height=0.8cm]{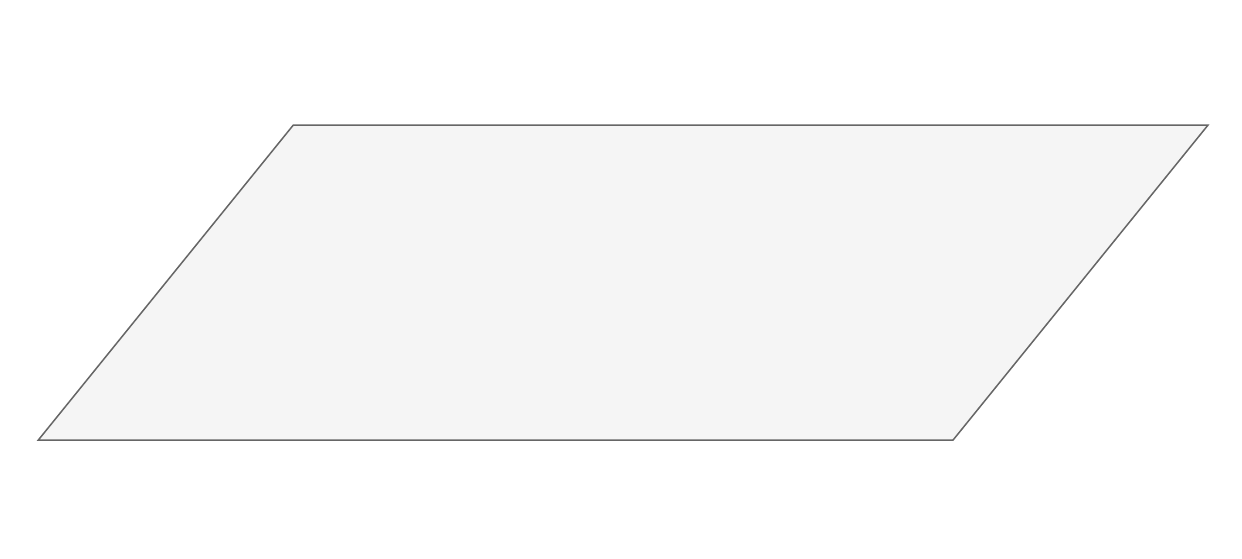}} \\
\hline
\textbf{Quality Goal} & A set of precise standards or criteria that must be met to achieve a broader goal. & 
\makebox[2cm][c]{\vspace{5pt}\includegraphics[height=0.8cm]{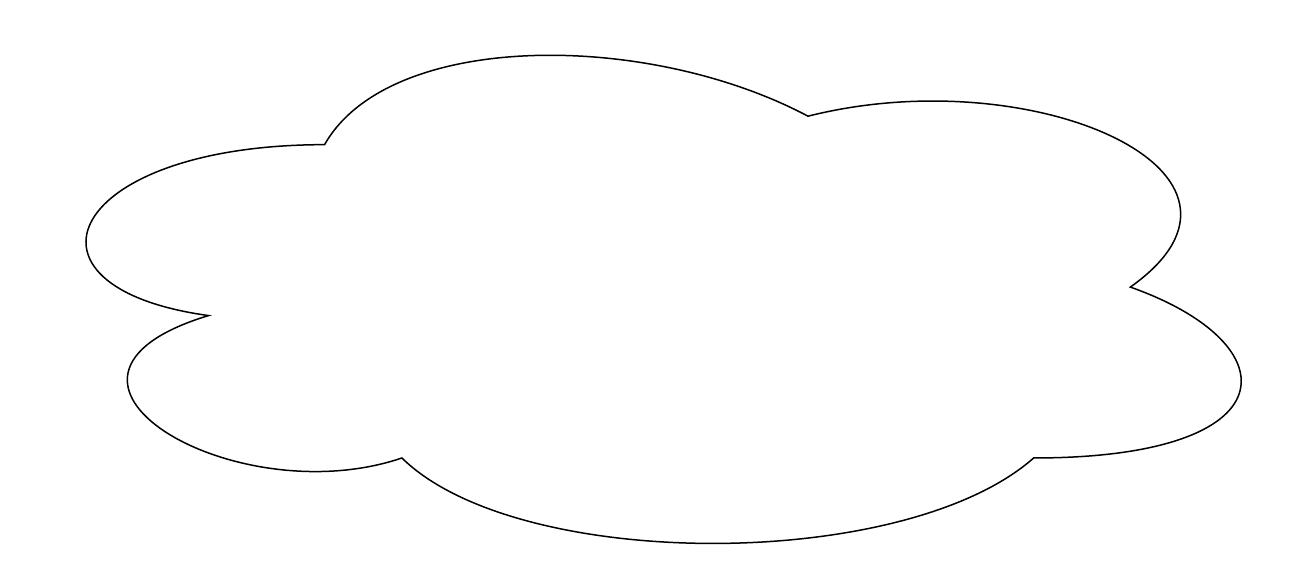}} \\
\hline
\textbf{Decomposition} & Relationship to illustrate the connection between two distinct goals, namely a higher level or parent goal and a lower level or sub-goal. & 
\makebox[2cm][c]{\vspace{0pt}\includegraphics[height=0.8cm]{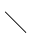}} \\
\hline
\textbf{Association} & Relationship used to connect a goal with other related elements that aid in defining or describing the goal. & 
\makebox[2cm][c]{\vspace{1pt}\includegraphics[height=0.8cm]{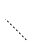}} \\
\hline
\textbf{Example} & A concrete instantiation used to illustrate a goal in the case study & 
\makebox[2cm][c]{\vspace{5pt}\includegraphics[height=0.8cm]{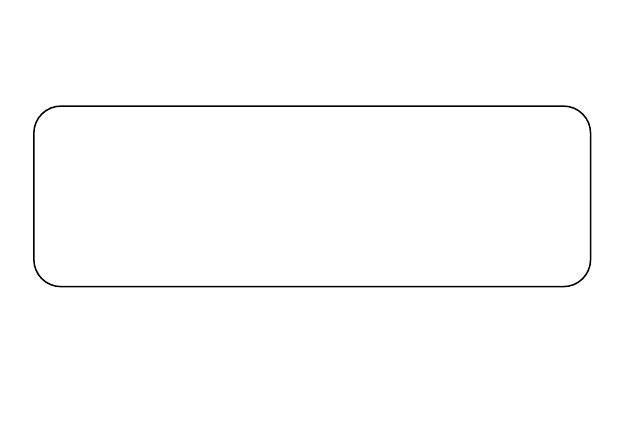}} \\
\hline
\end{tabular}

\caption{Goal Model Notation adopted in the use case.}
\label{table:goal_model_notation}
\end{table}

\subsection{Literature Review}\label{sec:literaturereview}

This literature review organizes prior work into four complementary streams. First, we discuss scholarly frameworks and taxonomies that conceptualize token economies at a theoretical level (Section~\ref{sec:theoreticalframeworks}). Second, we review special-purpose token-economy design approaches proposed for specific domains (Section~\ref{sec:specialpurpose}). Third, we examine literature addressing one of the three recurring token-economy design dimensions in isolation: governance (Section~\ref{sec:governance}), incentives (Section~\ref{sec:incentives}), and tokenomics (Section~\ref{sec:tokenomics}). Finally, we synthesize the resulting gaps and motivate the need for a stepwise, empirically grounded design method that integrates these dimensions.

\subsubsection{Theoretical Frameworks and Taxonomies of Token Economies}\label{sec:theoreticalframeworks}
The academic literature reports several theoretical frameworks that address token-economy design from different perspectives. Benedetti et al.~\cite{Benedetti2023} examine typical tokenomic designs, explore regulatory approaches, and review applications of utility tokens in domains such as decentralized finance and virtual reality platforms. Lage et al.~\cite{Lage2022} analyze fundamental characteristics of blockchain-based decentralized platform models. Khamisa~\cite{Khamisa2021} proposes a conceptual structure for token-economy design that includes the definition of economic goals, token design choices, and governance considerations.

\begin{figure}
   \hspace*{2cm}
    \centering
    \includegraphics[scale = 0.6]{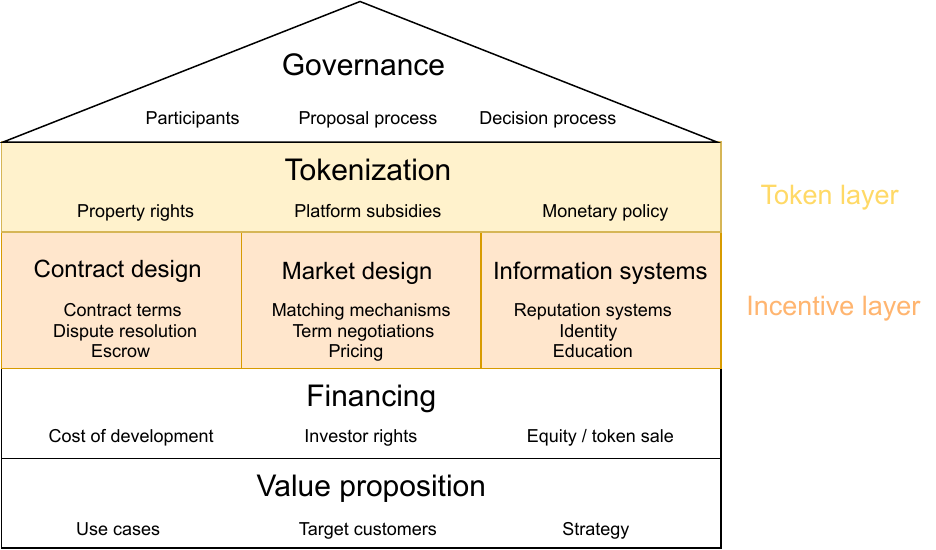}
    \caption{The House Framework~\cite{Barrera2020}.}
    \label{fig:house_framework}
\end{figure}

Barrera and Hurder~\cite{Barrera2020} introduce the \textit{House Framework} (Figure~\ref{fig:house_framework}), which decomposes the economic design of blockchain-based ecosystems into five layers: (1) value proposition, (2) financing, (3) incentives, (4) token design, and (5) governance. While the framework is useful for conceptual decomposition, it remains high-level and provides limited guidance on how to navigate among concrete design options within and across layers.

In addition to conceptual frameworks, several works focus on token classification and taxonomy. Freni et al.~\cite{Freni2022} map channels through which tokens can add value (technology, behavior, coordination) and highlight the relevance of token characteristics, monetary policy, and incentive mechanisms. Oliveira et al.~\cite{Oliviera2018} propose a guide to token design based on literature review and expert interviews. Tapscott~\cite{Tapscott2020} discusses token standards and a taxonomy-oriented perspective on tokens as digital assets. Overall, these contributions provide valuable constructs and categories, but they typically do not translate the constructs into an end-to-end, stepwise design method that integrates incentives, governance, and tokenomics while making decision points and trade-offs explicit.

\paragraph{Positioning relative to practitioner token-engineering and quantitative design frameworks}
In industry and adjacent research communities, token-engineering approaches often emphasize quantitative refinement, simulation, or state-space exploration. Examples include the simulation tooling and methodology developed by BlockScience\footnote{https://block.science/} and related work (e.g.,~\cite{liang2023cadcad,RiceZargham2019TokenEngineeringIII}), as well as state-space modeling frameworks for engineering crypto-economic systems (e.g.,~\cite{Zargham2018StateSpaceBlockchain,Zargham2020}). These approaches are particularly valuable for stress testing, parameter tuning, and exploring system dynamics once a candidate design is specified. However, they typically presuppose that the designer has already articulated (i) system goals, (ii) stakeholder roles, (iii) governance decision rights, and (iv) token issuance/allocation policies in a sufficiently structured way to be operationalized in models or simulations.
Against this background, the TEDM is positioned as a complementary, design-science artifact that supports the \emph{qualitative and structured derivation} of token-economy requirements and design propositions across incentives, governance, and tokenomics, thereby producing inputs that can subsequently be refined and validated quantitatively.

\subsubsection{Special Purpose Token Economy Design} \label{sec:specialpurpose}
A large part of token-design research focuses on special-purpose cases and domain-specific constraints, which limits transferability across domains. For instance, Wang et al.~\cite{Wang2020} and Dasaklis et al.~\cite{Dasaklis2019} focus on supply-chain contexts, while Hofmann et al.~\cite{Hofmann2021} address collaborative manufacturing. Hasan et al.~\cite{Hasan2018} and Zhang et al.~\cite{Zhang2019} propose domain-specific designs for the automotive sector. Direr et al.~\cite{Direr2022} develop an economic design for a Web3 game. Hou et al.~\cite{Hou2022} propose a mechanism for a voluntary public carbon market using a dual-token design. Kim et al.~\cite{Kim2021} propose a token-economic design for the Insolar business network. While these contributions are valuable within their domains, they typically do not provide a broadly applicable, stepwise method that integrates governance, incentives, and tokenomics for token economies with heterogeneous stakeholders and objectives.

\subsubsection{Governance}\label{sec:governance}
Several studies highlight the interconnection between token economies and DAOs, since token-based governance both relies on tokens (e.g., voting power and membership) and directly affects token supply, distribution, and incentive alignment. Avanzo et al.~\cite{avanzo2023modelling} emphasize the complexity of token-economy design in DAO settings and argue for modeling languages and methods that support governance grounded in sound token-economy design. Empirical studies examine governance in prominent DAOs, such as voting-power concentration and participation (e.g., Compound, Uniswap, ENS)~\cite{Fritsch2022}, and the impact of governance decisions on token economies (e.g., MakerDAO)~\cite{sun2021decentralization,zhao2022task}. Venugopalan et al.~\cite{venugopalan2023dance} consider the interconnection among governance, tokenomics, and incentives in DAOs, but the resulting method and tooling primarily target governance-related tensions and do not constitute an integrated, stepwise design method across all three pillars.

Additional work proposes governance models, evaluation perspectives, and analytical tools for blockchain governance. Bena and Zhang~\cite{Bena2022} study decentralized governance under heterogeneous contributor costs. Kiayas and Lazos~\cite{Kiayas2023} discuss evaluation of governance procedures. Liu et al.~\cite{liu2022} propose a governance framework including factors such as decentralization, authority, incentives, and accountability. Fernandez et al.~\cite{Fernandez2022} simulate voting-right concentration after token launches. Mohan et al.~\cite{Mohan2022} discuss challenges such as Sybil resistance and plutocracy. Wang et al.~\cite{wang2023blockchain} propose an ontology of blockchain governance that includes incentive-related concepts. Overall, this stream provides important constructs and empirical insights, but practical stepwise guidance that also incorporates tokenomics decisions (issuance schedules, allocation strategies, price-management mechanisms) remains limited.

\subsubsection{Incentive Structures}\label{sec:incentives}
A substantial body of work addresses incentives in blockchain systems, including surveys of incentive mechanisms~\cite{wang2019survey,yu2018survey,huang2019survey,han2022can}. However, many contributions focus primarily on mining and base-layer protocol incentives rather than application-level token economies. Complementary work examines how token design shapes participant incentives in ledger-based ecosystems, sometimes in domain-specific contexts (e.g., supply chains or data markets)~\cite{Jürjens2021}. Others propose incentive models aimed at improving business-market profitability~\cite{Guo2022} or analyze token incentives in competitive platform settings (e.g., decentralized exchanges)~\cite{Liu2022TokenIncentives}. DAO-oriented models frequently incorporate incentives as a core component~\cite{qin2022web3,wang2019decentralized} but typically provide limited stepwise design guidance and often do not integrate tokenomics decisions such as issuance, allocation, and price-management policies.

\subsubsection{Tokenomics}\label{sec:tokenomics}
Tokenomics is frequently studied under the lenses of token monetary policy, supply strategies, and distribution mechanisms. Prior work examines airdrops and strategic behaviors (including Sybil participation)~\cite{Lommers2023,Liu2022FightingSybils}, relates tokenomics to concepts in finance (e.g., shares, profits, dividends)~\cite{Carvalho2022}, and proposes valuation perspectives for DAOs~\cite{Lommers2022}. Other studies focus on ICO design and issuance caps~\cite{Gan2022}, as well as token launch practices and valuation methods for digital assets~\cite{Kaal2022}. Additional work analyzes wealth concentration~\cite{Kusmierz2022, 10.1145/3589335.3651481, eisermann2025concentration} and mechanisms such as burns and buybacks through a corporate-finance lens~\cite{Allen2022}.

Among the closest stepwise contributions, Kim and Chung~\cite{Kim2018} analyze the tokenized social network Steemit and propose an eight-step design flow for a successful ICO. While valuable, this flow is tailored to an ICO setting and does not address governance-property specification and incentive-structure design in an integrated manner.

\paragraph{Synthesis and gap}
Across the reviewed streams, prior work provides (i) conceptual frameworks and taxonomies that define relevant constructs, (ii) domain-specific token designs that incorporate local constraints, and (iii) pillar-specific insights on governance, incentives, or tokenomics. Nevertheless, empirically grounded, stepwise guidance that integrates the three pillars into a coherent design method, while making decision points, trade-offs, and boundary conditions explicit, remains limited. This motivates the development of the Token Economy Design Method (TEDM) as a DSR artifact that consolidates and operationalizes design propositions across incentives, governance, and tokenomics, and evaluates them formatively through an empirical design case and expert feedback.

\subsection{Use Case - Currynomics Ecosystem}\label{sec:usecase} 
In this study, the TEDM is formatively evaluated using the Currynomics ecosystem as an empirical design case. The case is suitable because it entails an integrated token-economy design problem in which incentives, governance, and tokenomics must be specified jointly under real-world constraints (e.g., asset backing, organizational roles, and stakeholder trust). At the same time, we explicitly treat Currynomics as a bounded problem instance: it does not represent all classes of token economies (e.g., governance-minimal protocols or purely algorithmic stablecoins), and the case is used to ground and refine the method rather than to claim statistical generalization.

Currynomics is a decentralized blockchain ecosystem that links the value of its stablecoin (the Redcurry token) to the net asset value (NAV) of a commercial real estate (CRE) portfolio~\cite{RedCurry2023}. The design of Redcurry is motivated by a common challenge among stablecoins: token holders often cannot fully verify or trust the validity and availability of the underlying collateral, which increases perceived shortfall risk and threatens the peg~\cite{hafner2024four}. In Currynomics, Redcurry represents the value of the CRE portfolio and is designed to maintain the peg through an asset-backed mechanism~\cite{RedCurry2023}.

By functioning as a means of payment and a store of value, Redcurry resembles established stablecoins such as USDT, but differs in that its supply is not fixed~\cite{RedCurry2023}. Moreover, Redcurry is not a pure asset token: holders do not obtain direct legal rights to specific underlying properties. Instead, the ecosystem reinvests financial gains from the portfolio back into the system. In this sense, the design aims to ``commodify'' real estate by creating a bridge through which capital can move from the traditional economy to the crypto economy while maintaining an asset-backed value anchor.

\begin{figure}
    \centering
    \includegraphics[scale = 0.8]{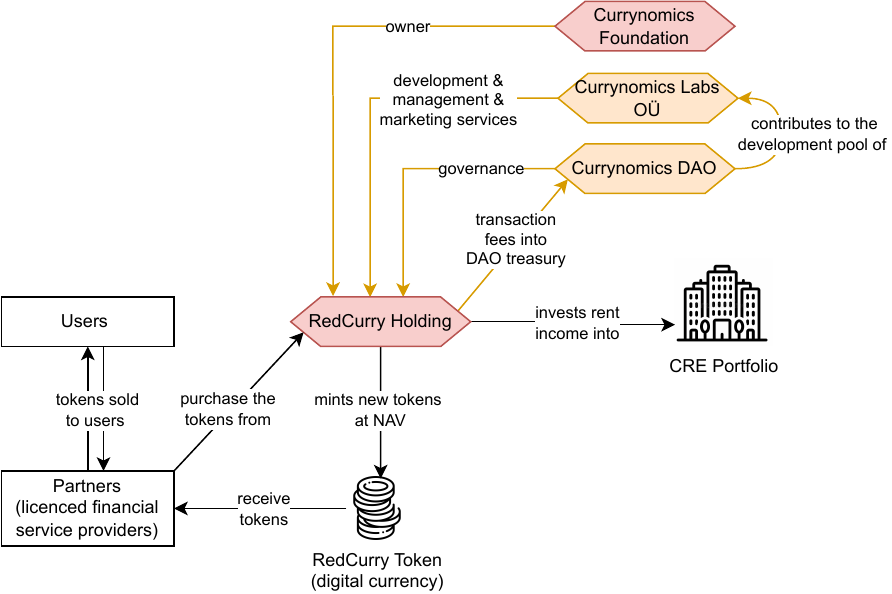}
    \caption{Stakeholder map of the Currynomics ecosystem.}
    \label{fig:stakeholdermap}
\end{figure}

The Currynomics ecosystem is summarized in Figure~\ref{fig:stakeholdermap}. The core bodies of the ecosystem are marked in red: Redcurry Holding mints Redcurry tokens, acquires real estate in the CRE portfolio, and distributes tokens to partners. Marked in orange are the developers and maintainers of the ecosystem: Currynomics Labs O\"U provides development, marketing, and management services, while Currynomics DAO is the governing body of the token economy and uses DAO tokens in its decision-making operations. Currynomics DAO operates as a decentralized organization that is distinct from the other legal entities in the ecosystem.

The token economy supporting Redcurry faces multiple challenges originating from the surrounding business environment. A key challenge is to identify the levers that build trust for users (Redcurry holders) to purchase and retain the token in the long term, given the availability of competing investment alternatives outside the crypto economy (e.g., low-risk real-estate funds). Trust is tightly connected to governance design, raising questions about whether (and how) community participation can be sustained without exposing the system to governance capture or misaligned incentives. Finally, the introduction of a governance token (DAO token) requires careful tokenomics decisions concerning issuance timing, allocation, and distribution mechanisms, as these choices shape both governance power and long-term incentive alignment.

\section{Token Economy Design Method} \label{sec:tokeneconomydesignmodel}
The token-economy design aspects introduced in Sections~\ref{sec:governance}, \ref{sec:incentives}, and \ref{sec:tokenomics} are consolidated into the Token Economy Design Method (TEDM), which we describe in this section.

TEDM does not follow a predominantly quantitative approach (e.g., optimization or simulation-based search), as commonly emphasized in parts of the practitioner token-engineering literature. Instead, TEDM emerged through an iterative synthesis of relevant scholarly literature and formative evaluation through an empirical design case. This approach deliberately prioritizes actionable relevance and interpretability over formal completeness; accordingly, TEDM should be understood as a set of empirically grounded design propositions that make decision points, trade-offs, and risks explicit, rather than as a predictive model of token-economy outcomes.

\begin{figure}
    \centering
    \includegraphics[width = 0.92\linewidth]{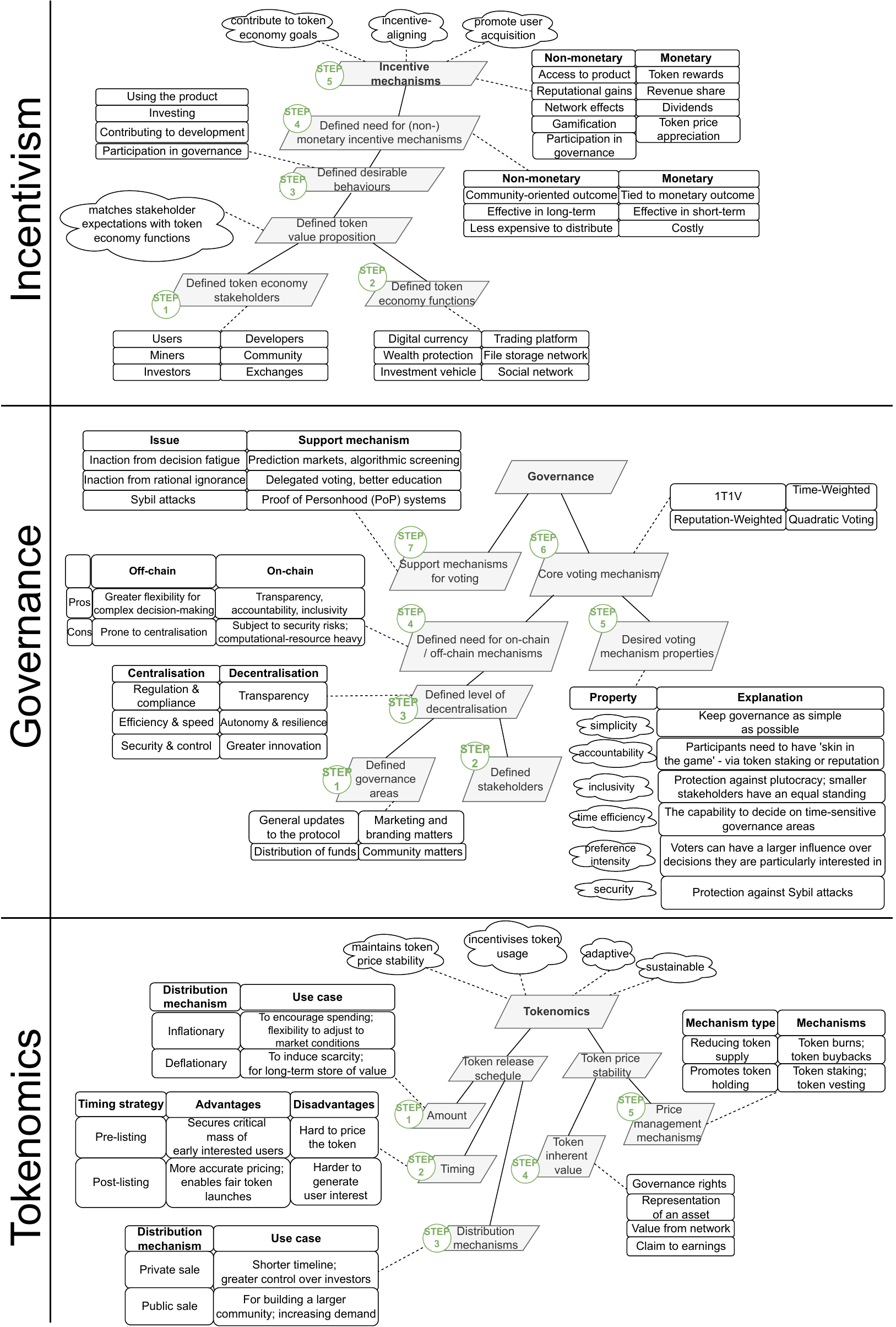}
    \caption{Token Economy Design Method (TEDM).}
    \label{fig:tokeneconomydesignmodel}
\end{figure}

TEDM is applied to the Currynomics case introduced in Section~\ref{sec:usecase}. The results are evaluated through a case-based demonstration and expert interviews (Section~\ref{sec:evaluation}). Figure~\ref{fig:tokeneconomydesignmodel} summarizes the three sequential method components and their stepwise structure.

TEDM consists of three components: incentive structures, governance, and tokenomics. Each component is detailed in the subsections that follow.

\subsection{Incentives} \label{sec:TEDMincentives}

Well-designed incentive structures are necessary to align stakeholder behavior with the token economy's goals. As discussed by Yoo et al., it is crucial to understand which user groups participate in the token economy and what benefits they receive from reward and penalty mechanisms~\cite{Yoo2021}. According to~\cite{Khamisa2021}, key principles for incentive design are to encourage actions that advance the primary objectives of the token economy, discourage destructive behavior, and account for heterogeneous stakeholder needs. Incentives are also relevant because they act as remuneration mechanisms that can support the long-term viability of the ecosystem~\cite{Lee2019}.

This subsection responds to RQ1 (Section~\ref{introduction}) by providing stepwise guidance for establishing incentive mechanisms. In TEDM, each step is formulated as a design proposition: it clarifies the design goal addressed, the rationale for the step, and the expected output artifact thereby making trade-offs and risks explicit rather than providing a purely procedural checklist.

\begin{table}[ht]
    \centering
    \begin{tabular}{|p{9cm}|p{6cm}|}
        \hline
        \textbf{Incentive Design Goals} & \textbf{TEDM Incentive design} \\
        \hline
        \multirow{2}{9cm}{ Defining the token economy value proposition} &1) Identify the token economy stakeholders \\
        \cline{2-2}
         &2) Define the token economy functions \\
        \hline
         Identifying behaviors that contribute to system utility &3) Determine desirable behaviors \\
        \hline
        \multirow{2}{9cm}{Selecting incentive mechanisms to promote desirable behaviors} &4) Define the type of incentive mechanism. \\
        \cline{2-2}
         &5) Specify incentive mechanisms \\
        \hline
    \end{tabular}
    \caption{Incentive design goals and corresponding TEDM steps.}
    \label{tab:areas}
\end{table}

Steps 1 and 2 define the token economy value proposition in terms of (i) actors and (ii) functions the system is expected to perform. After the value proposition is established, Step 3 identifies desirable behaviors that support system utility. Steps 4 and 5 then determine the type of incentives required and specify concrete incentive mechanisms. Each step is defined below, followed by its application to the Currynomics case.

\subsubsection{Step 1 - Identify the stakeholders}
\label{stakeholders}

Identifying stakeholder categories and their motivations is a key prerequisite for incentive design~\cite{Khamisa2021}. Prior work proposes different stakeholder taxonomies. Davidson~\cite{Davidson2021} distinguishes maintainers, contributors, and users. Allen and Berg~\cite{Allen2020} include token holders, developers, founding teams, miners/validators, and indirect stakeholders such as government and venture capital. Voshmgir~\cite{Voshmgir2020} highlights nodes, developers, miners, and market actors (e.g., market makers). Liu et al.~\cite{liu2022} distinguish project teams, node operators, users, application providers, and regulators. A single individual may belong to multiple stakeholder groups simultaneously.

In the Currynomics case, we identify:
\begin{itemize}
    \item \textbf{Users}: Redcurry token holders who purchase and hold the token primarily for financial purposes.
    \item \textbf{Partners}: licensed financial institutions that purchase tokens from Redcurry Holding and distribute them to end users for a transaction fee.
    \item \textbf{Developers}: Currynomics Labs O\"U, providing development, marketing, and management services for the ecosystem.
    \item \textbf{Community members}: holders of the Currynomics governance token (DAO token) participating in governance processes.
    \item \textbf{Investors}: actors providing capital either by purchasing/holding governance tokens or by investing in Currynomics Labs O\"U to support ecosystem development.
\end{itemize}

\subsubsection{Step 2 -  Identify the functions of the token economy.} 

Blockchain ecosystems connect stakeholders to create value based on shared resources~\cite{Barrera2020}. Value creation can include the transfer of goods, means of payment, monitoring of asset status, enabling access to a service/product, or voting~\cite{Schubert2021}. Sockin and Xiong~\cite{Sockin2020} argue that value may stem from (i) enabling transactions among a large set of users or (ii) providing an investment vehicle. From a financial perspective, systems may also aim at price stability (as common in stablecoins) or wealth protection against inflation~\cite{Kampakis2022}.

The core objectives of the Currynomics ecosystem are predominantly financial. The Redcurry token, minted against real-estate acquisition, is intended to support:
\begin{itemize}
    \item wealth protection and liquidity parking for savings;
    \item an investment vehicle linked to CRE portfolio value appreciation;
    \item a means of payment;
    \item collateral utility in credit settings enabled by price stability.
\end{itemize}
 
\subsubsection{Step 3 - Define the desirable behaviors}  
In an effective token economy, participants' actions should support system utility rather than only individual short-term gains~\cite{Khamisa2021,Pazaitis2017}. Khamisa~\cite{Khamisa2021} refers to this as \textit{token--network fit}. Yoo~\cite{Yoo2021} argues that desirable behaviors are determined by whether actions add long-term economic value to the ecosystem. Long-term participation and retention are frequently highlighted as particularly valuable behaviors~\cite{Yoo2021,Kang2019}.

In Currynomics, the following behaviors were identified:
\begin{itemize}
    \item \textbf{Users}: hold Redcurry as a long-term store of value and use it without persistently selling below its reference value/market price anchor.
    \item \textbf{Community members}: participate in discussion and decision-making via the DAO token.
    \item \textbf{Investors}: provide development liquidity (e.g., by purchasing/holding the DAO token or investing in the development entity).
\end{itemize}

While Redcurry can support multiple financial functions, its primary intended use in the early phases is long-term holding to build stability and trust. Trading can be neutral in later phases provided it does not systematically undermine the price anchor during low-liquidity periods.

\subsubsection{Step 4 - Select incentive-mechanism types } 

Incentives can be monetary or non-monetary. Monetary incentives include \textit{token rewards} for actions~\cite{liu2022}, \textit{staking} mechanisms that lock tokens to support ecosystem operations~\cite{Freni2022,Lommers2022}, \textit{liquidity mining} by providing liquidity to DEX/DeFi protocols~\cite{Liu2022TokenIncentives}, and \textit{growth expectations} concerning platform success that motivate participation without direct token earnings~\cite{Freni2022,Drasch2020}. Non-monetary incentives discussed in the literature include access to services, reputation mechanisms, and governance participation~\cite{Freni2022}, as well as network effects and gamification~\cite{Kim2018}.

\begin{table}[h]
\centering
\begin{tabularx}{\textwidth}{|l|X|}
\hline
\textbf{Incentive Mechanism Type} & \textbf{Mechanism Description} \\
\hline
\multirow{4}{*}{Monetary} & Token Rewards (User actions)~\cite{liu2022} \\
\cline{2-2}
                          & Staking (Token lock-up for support)~\cite{Freni2022,Lommers2022} \\
\cline{2-2}
                          & Liquidity Mining (Earning via DEX/DeFi)~\cite{Liu2022TokenIncentives} \\
\cline{2-2}
                          & Growth Expectations (platform success)~\cite{Freni2022,Drasch2020} \\
\hline
\multirow{3}{*}{Non-Monetary} & Access to services~\cite{Freni2022} \\
\cline{2-2}
                             & Reputation mechanisms (enhanced reputation)~\cite{Freni2022} \\
\cline{2-2}
                             & Governance Participation (decision-making)~\cite{Freni2022} \\
\cline{2-2}
                             & Network Effects~\cite{Kim2018} \\
\cline{2-2}
                             & Gamification~\cite{Kim2018} \\
\hline
\end{tabularx}
\caption{Monetary and non-monetary incentive-mechanism types.}
\label{tab:incentive_types}
\end{table}

\subsubsection{Step 5 - Specify incentive mechanisms}  
Concrete incentive mechanisms should (i) contribute to token-economy goals, (ii) motivate behavior that supports system utility, and (iii) attract and retain participants. These can be treated as quality goals, i.e., criteria that incentive mechanisms should satisfy.

In Currynomics, Redcurry Users are primarily motivated by monetary incentives, in particular gains associated with Redcurry value appreciation and stability. By contrast, the DAO token primarily aims to incentivize governance participation and coordinated decision-making. Accordingly, non-monetary incentives such as reputational gains and the intrinsic value of governance participation are particularly relevant for community members, while the potential appreciation of the DAO token can serve as an additional monetary incentive for investors.

\subsection{Governance of a Token Economy}\label{sec:TEDMgovernance}
Figure~\ref{fig:tokeneconomydesignmodel} summarizes seven governance-design steps in TEDM. The goal is to structure governance decisions according to project requirements while making explicit the main trade-offs and risks (e.g., plutocracy/governance capture, Sybil attacks, low participation, and decision fatigue)~\cite{ma2025understanding}.

\subsubsection{Step 1 - Define governance areas} 
Governance areas depend on the system goals and the scope of what the community is allowed (and able) to decide. Governance decisions can include protocol upgrades and parameter changes, ecosystem service/product development, recruitment and management, treasury and token-asset management, and changes to governance rules themselves~\cite{Khamisa2021}. Governance may also include emergency actions (e.g., triggering shutdown modes) and infrastructure funding decisions~\cite{Bersani2022}, as well as compensation and resource-allocation policies~\cite{Barrera2020}.

In Currynomics, we identify four broad governance areas:
\begin{itemize}
    \item \textbf{Treasury management}: allocation and management of the DAO treasury funds.
    \item \textbf{Governance and process}: governance structure, voting procedures, community goals, and community sentiment toward proposals.
    \item \textbf{Protocol upgrades}: implementation of upgrades, adjustment of global and treasury-specific parameters, and replacement of modular smart contracts.
    \item \textbf{Tokenomics}: decisions related to the issuance, allocation, and distribution of governance tokens and other token-policy parameters.
\end{itemize}

\subsubsection{Step 2 - Define governance stakeholders and roles} 
For each governance area, the relevant stakeholders and their roles should be identified (e.g., who proposes, who deliberates, who votes, who executes, and who is accountable). In Currynomics, the stakeholder groups are defined in Section~\ref{stakeholders}; here, TEDM uses these categories to assign decision rights and responsibilities across the governance areas.

\subsubsection{Step 3 - Define the target level of decentralization}

At this stage, the desired degree of decentralization should be determined. Liu et al.~\cite{liu2022} distinguish (i) \textit{private and centralized} ecosystems (e.g., enterprise blockchains), (ii) \textit{public but centralized} token economies where governance is formally open but effectively controlled by privileged actors, and (iii) \textit{public and decentralized} systems without privileged stakeholders, enabling broad participation.

To operationalize decentralization targets, public token economies are commonly assessed using distribution-based indicators such as the \textit{Gini coefficient} and the \textit{Nakamoto coefficient}~\cite{feichtinger2023hidden, eisermann2025concentration}. The Gini coefficient quantifies inequality in a resource distribution (e.g., voting power), ranging from \(0\) (perfect equality) to \(1\) (maximal concentration):

\begin{equation}
G = \frac{\sum_{i=1}^{n} \sum_{j=1}^{n} |x_i - x_j|}{2n^2 \bar{x}} .
\end{equation}

A higher Gini value indicates stronger concentration and, therefore, greater centralization.

The Nakamoto coefficient captures how many independent entities are required to control more than 50\% of a resource (e.g., voting power):

\begin{equation}
N = \min \left\{ k \;\middle|\; \sum_{i=1}^{k} S_i > \frac{1}{2} \sum_{j=1}^{n} S_j \right\} .
\end{equation}

Lower Nakamoto values indicate that few entities can form a controlling coalition (more centralization), while higher values indicate broader dispersion of control.

These indicators can support (i) setting decentralization targets at design time and (ii) monitoring whether the implemented governance remains aligned with those targets once the system is operational. In practice, decentralization is multi-dimensional (it may differ across token ownership, voting participation, proposer power, and execution privileges), so these metrics should be interpreted as informative signals rather than exhaustive measures.

In Currynomics, some governance decisions remain necessarily centralized due to legal and operational constraints. For instance, Redcurry issuance is governed by code, while the reserve-asset composition is managed by an investment board. The composition of the Currynomics Foundation board, which oversees ecosystem integrity, is also not subject to community votes. However, multiple governance areas remain open to decentralized participation, allowing community members to influence decisions while maintaining essential oversight.

\subsubsection{Step 4 - Define the need for on-chain and off-chain mechanisms} \label{onchainoffchain}
On-chain governance refers to processes where proposals and voting are executed through the protocol (e.g., via smart contracts), potentially enabling automated enactment of accepted changes~\cite{Reijers2018}. Off-chain governance refers to deliberation and decision-making processes that occur outside the protocol (e.g., forums, social channels), often used for discussion, signaling, and agenda setting~\cite{Allen2022}.

Currynomics aims to keep governance as transparent and inclusive as possible; therefore, most governance areas are decided via on-chain voting mechanisms. However, some stages are intentionally off-chain, such as proposal filtering and early deliberation (``temperature checks'') in a community forum. If a proposal passes the temperature check, it can be escalated to a formal vote. This hybrid structure balances transparency and inclusivity with practical deliberation needs that are difficult to accommodate purely on-chain.

\subsubsection{Step 5 - Define desired voting-mechanism properties}
No single voting mechanism is universally optimal across criteria; selection involves trade-offs~\cite{Kiayas2023}. In TEDM, candidate mechanisms (Step~6) are assessed against the following desired properties:
\begin{itemize}
    \item \textbf{Simplicity}: procedures should be understandable and implementable without unnecessary complexity.
    \item \textbf{Accountability}: participants should have ``skin in the game'' (e.g., committed tokens, time, or reputation) to promote responsible decisions~\cite{Kiayas2023}.
    \item \textbf{Inclusivity}: mechanisms should avoid systematically marginalizing small stakeholders purely due to limited financial means~\cite{Mohan2022}.
    \item \textbf{Time efficiency}: urgent decisions should be feasible within time constraints~\cite{Kiayas2023}.
    \item \textbf{Intensity of preferences}: mechanisms should allow expressing how strongly voters care about an issue~\cite{Mohan2022}.
    \item \textbf{Security}: mechanisms should mitigate voter fraud (e.g., Sybil attacks)~\cite{Mohan2022} and reduce centralization/capture risks associated with concentrated voting power~\cite{Buterin2021}.
\end{itemize}

\subsubsection{Step 6 - Core voting mechanisms} \label{corevotingmech}
We distinguish four core voting-mechanism families discussed in the literature:
\begin{itemize}
    \item \textbf{1-token-1-vote (1t1v)} assigns voting power proportional to token holdings~\cite{Mohan2022}. It is simple and time-efficient, but can enable vote buying and plutocracy, reducing inclusivity and increasing capture risk~\cite{Buterin2021}.
    \item \textbf{Time-weighted voting} introduces a time commitment to increase accountability and mitigate short-term capture dynamics~\cite{Mohan2022}. Time-weighting can improve accountability/security compared to pure 1t1v, but often reduces time efficiency and increases complexity. Common variants include:
    \begin{itemize}
        \item \textbf{Conviction voting}: sustained support increases ``conviction'' for a proposal over time~\cite{FaqirRhazoui2021}.
        \item \textbf{Vote-escrow (ve-token)}: users lock tokens for a period to receive increased voting power~\cite{Mohan2022}.
        \item \textbf{Bond voting}: voting power increases with staked resources via bond-like commitments~\cite{Mohan2022}.
    \end{itemize}
    \item \textbf{Reputation-weighted voting} weights votes by reputation rather than token wealth~\cite{ValienteBlazquez2021}. This can improve inclusivity, but requires additional mechanisms to quantify and update reputation and can introduce new attack surfaces and governance overhead. It is used, for example, in Colony~\cite{Mohan2022}.
    \item \textbf{Quadratic voting} allocates voice credits such that the marginal cost of additional votes increases quadratically, aiming to reflect intensity of preferences~\cite{WrightJr2019,Dimitri2022,benhaim2025balancing}. It can improve inclusivity relative to 1t1v, but its real-world robustness and resistance to strategic behavior remain debated~\cite{Dimitri2022}.
\end{itemize}
    
The mechanisms are mapped to the desired properties in Table~\ref{tab:voting_matrix}. In the table, a full node indicates strong fulfillment, a half-filled node indicates partial fulfillment, and an empty node indicates weak fulfillment based on the cited literature.
\begin{table}
    \centering
    \includegraphics[width=0.7\linewidth]{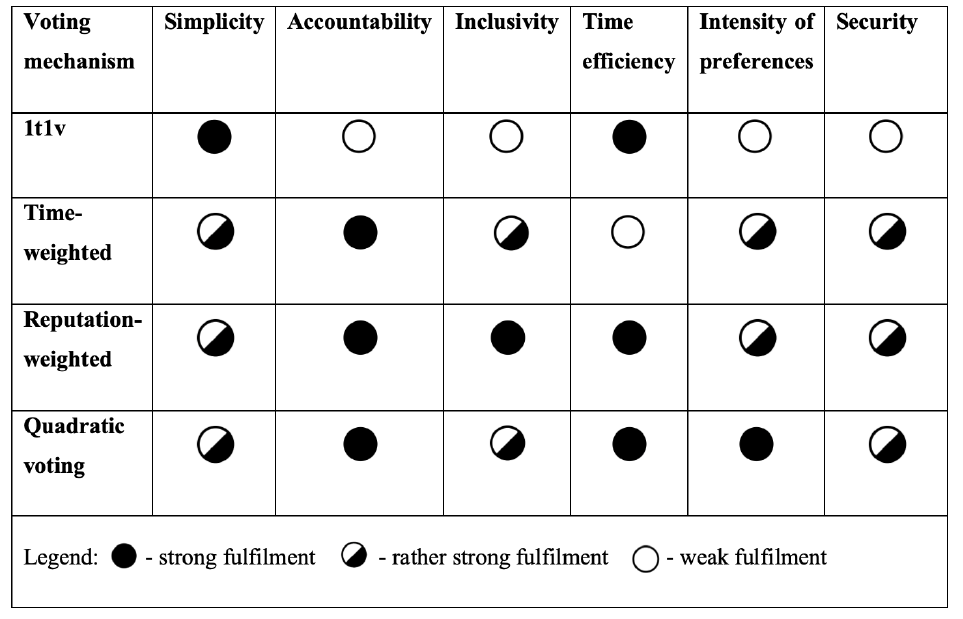}
    \caption{Voting-mechanism families mapped to desired properties, with scores derived from the referenced literature.}
    \label{tab:voting_matrix}
\end{table}

Achieving both Sybil resistance and plutocracy resistance simultaneously is challenging when voting power depends predominantly on a single factor (e.g., staked tokens)~\cite{Mohan2022}. Including time or reputation can mitigate certain risks, but does not eliminate Sybil behavior or capture incentives entirely; therefore, no mechanism achieves a perfect score in the \textit{security} dimension.

In Currynomics, the DAO prioritizes \textit{accountability} and \textit{security} due to the system's financial focus. The selection process rules out 1t1v because of its vulnerability to plutocracy and capture dynamics. Among remaining options, Currynomics considered implementation and communication complexity as a key constraint. Reputation-weighted voting was deemed unsuitable because it requires additional reputation-assessment infrastructure. Quadratic voting was considered not worth the added complexity for the anticipated governance needs. Consequently, Currynomics selected \textit{time-weighted voting}, and within that family, \textit{conviction voting} as a comparatively simple option. This choice is case-dependent: TEDM provides a structured rationale for selecting a mechanism under given constraints rather than claiming that a single mechanism generalizes to all token economies.

\subsubsection{Step 7 - Support mechanisms for Voting}\label{supportmechanism}
As token economies mature, governance can face issues that core voting mechanisms do not directly address. TEDM therefore includes optional support mechanisms that can be introduced when specific needs arise:
\begin{itemize}
    \item \textit{Decision fatigue / proposal overload}: too many proposals can reduce participation and decision quality. Mitigations include agenda-setting, proposal pre-screening, prediction markets, or algorithmic filtering~\cite{Ding2021,Barrera2020}.
    \item \textit{Rational ignorance}: participants may avoid information gathering when its costs exceed perceived benefits~\cite{Kiayas2023}. Mitigations include delegated voting, improved information design, and structured deliberation~\cite{weidener2025delegated}.
    \item \textit{Sybil attacks}: identity splitting can occur even when voters commit tokens, time, or reputation. Mitigations include identity systems such as Proof of Personhood (PoP), acknowledging potential trade-offs in privacy and accessibility~\cite{Mohan2022}.
\end{itemize}

The Currynomics team plans to adopt a simple mechanism initially to support early community engagement, and to introduce more advanced support mechanisms only as governance complexity and participation needs increase.

\subsection{Tokenomics}\label{sec:TEDMtokenomics}
This subsection responds to RQ3 by providing stepwise guidance for tokenomics design in TEDM. In the method, tokenomics is treated as (i) a \emph{release and distribution policy} (Steps 1--3) and (ii) a \emph{value and price-management policy} (Steps 4--5). Each step is formulated as a design proposition that clarifies the design goal, typical options discussed in the literature, and the expected output artifact (e.g., a supply policy, a timing plan, or a price-management toolbox). Figure~\ref{fig:tokeneconomydesignmodel} summarizes these five steps.

\subsubsection{Step 1 - Define the token supply policy}
A first tokenomics decision concerns whether supply is capped or uncapped, and how minting and burning affect circulating supply over time. Prior work often distinguishes \emph{uncapped} (inflationary) models, where supply can expand over time, from \emph{capped} (deflationary or fixed-supply) models, where an upper bound exists~\cite{Kaal2018,Gan2020,Kaal2022a}. A simple accounting identity for token supply over discrete time steps is:

\[
S_t = \min\left(S_{\max}, S_{t-1} + M_t - B_t\right)
\]

where \(S_t\) is circulating supply at time \(t\), \(M_t\) newly minted tokens, and \(B_t\) burned tokens. For capped designs, an additional constraint \(S_t \le S_{\max}\) is imposed.

Furthermore, the constraints imposed below are specific to inflationary or deflationary issuance models.

\begin{itemize}
    \item \textbf{Inflationary Model:}
    \begin{itemize}
        \item \( S_{\max} \) is undefined (no cap).
        \item \( M_t > B_t \) (net supply increases over time).
    \end{itemize}
    
    \item \textbf{Deflationary Model:}
    \begin{itemize}
        \item \( S_{\max} \) is fixed (supply cannot exceed cap).
    \end{itemize}
\end{itemize}

\textit{Design implication and trade-off.} Capped supplies can support scarcity narratives and long-term holding, but may reduce spending willingness for governance or utility purposes if the token is primarily treated as a store of value. Uncapped supplies can support flexibility and responsiveness to demand but may increase concerns about dilution and long-run credibility if minting rules are not transparent and constrained.

\textit{Currynomics.} Redcurry issuance is demand-driven and asset-backed: new Redcurry tokens are minted as the underlying CRE portfolio expands. This resembles an uncapped mechanism in operational terms, but the minting rule is constrained by the collateral base rather than discretionary emission. By contrast, the Currynomics DAO token follows a capped supply policy. The team acknowledges the trade-off that strong scarcity incentives can reduce willingness to deploy DAO tokens for governance-related participation; the governance design (e.g., token locking for voting) can partially mitigate this by decoupling participation incentives from pure ``spend vs. hold'' dynamics.

\subsubsection{Step 2 - Define the timing strategy (pre-launch vs. post-launch)}
Token releases can occur \emph{pre-launch} (before product/protocol launch) or \emph{post-launch} (after launch)~\cite{Fritsch2022}. Pre-launch issuance can fund development and bootstrap early stakeholders (often via escrow/vesting structures)~\cite{Kaal2018}. Post-launch issuance can follow a schedule or be demand-driven; demand-driven issuance can help align supply expansion with adoption, which may reduce certain volatility pressures compared to large upfront releases~\cite{Kaal2018}.
\textit{Design implication and trade-off.} Pre-launch issuance improves early financing and coordination but increases risks of misalignment (e.g., early concentration, opportunistic selling) if vesting and lockups are weak. Post-launch issuance can better reflect realized demand but may slow early ecosystem growth if participation incentives require upfront distribution.

\textit{Currynomics.} Redcurry cannot be meaningfully issued pre-launch because issuance requires acquisition and management of the underlying real-estate portfolio; hence, it is effectively post-launch and collateral-driven. The DAO token can be issued partly pre-launch to secure early contributors and bootstrap governance capacity, with controls (e.g., vesting) to limit early sell pressure.

\subsubsection{Step 3 - Define the distribution mechanism (public vs. private channels)} \label{tokendistribution}
Token distribution can occur through different channels, including:
\begin{itemize}
    \item \textbf{Private sale}: selling tokens to a limited set of investors (often discounted)~\cite{Howell2020}.
    \item \textbf{Public sale}: selling tokens broadly (e.g., ICO/IEO or other public offerings)~\cite{Gan2022,Freni2022,Lyandres2022}.
    \item \textbf{Airdrops}: distributing tokens to users for free to promote adoption~\cite{Lommers2023,Liu2022FightingSybils,allen2024crypto}.
\end{itemize}

\textit{Design implication and trade-off.} Private sales can accelerate funding and attract strategic partners but may increase concentration and perceived unfairness. Public sales can broaden participation and signal demand but face regulatory and market-manipulation risks. Airdrops can bootstrap adoption but may allocate voting power to recipients with low governance competence or low willingness to participate, potentially harming governance quality~\cite{Davidson2021}.

\textit{Currynomics.} Redcurry distribution is structured as a public channel via licensed financial institutions (partners), consistent with its asset-backed nature and compliance constraints. The DAO token is distributed partly through a pre-launch private sale (to founders, early contributors, advisors, and early-stage investors) and partly through post-launch public mechanisms. The remaining pool is reserved for liquidity provision and ecosystem incentives (e.g., bug bounties, community management, marketing), with an additional reserve for future ecosystem needs.

\subsubsection{Step 4 - Define value-capture mechanisms (``inherent value'')} \label{governancevalue}
Token price sustainability depends on credible value-capture mechanisms and on how token demand relates to system utility. While some authors argue that tokens lack intrinsic value in a strict sense~\cite{Allen2022,Kaal2018}, the literature discusses recurring channels through which tokens can capture value:
\begin{itemize}
    \item \textbf{Governance rights}: participation in decision-making~\cite{Carvalho2022,Lommers2022}.
    \item \textbf{Claims linked to assets}: representation tied to real-world or on-chain assets (e.g., collateral-backed structures)~\cite{Freni2022,Kaal2018}.
    \item \textbf{Network value}: value influenced by adoption, trust, and utility of the ecosystem~\cite{Freni2022,Kaal2018}.
    \item \textbf{Claims on earnings}: structures where token holders benefit from fee flows or economic surplus (directly or indirectly)~\cite{Freni2022,Kaal2022a}.
\end{itemize}

\subsubsection{Step 5 - Define price-management mechanisms (stability toolbox)}
Designers often introduce mechanisms to mitigate excessive volatility, preserve trust, and manage adverse scenarios (e.g., speculative pressure, liquidity shocks, or rapid sell-offs). Prior work highlights that speculation can crowd out utility users under high volatility, undermining long-run adoption~\cite{Li2018,Mayer2019}. Common mechanisms discussed in the literature include:
\begin{itemize}
    \item \textbf{Token burns}: reducing circulating supply by permanently removing tokens~\cite{Allen2022}.
    \item \textbf{Staking/locking}: reducing circulating supply and aligning incentives through lock-up commitments~\cite{Lommers2022}.
    \item \textbf{Buybacks}: purchasing tokens from the market using treasury or revenue, reducing circulating supply and signaling resources~\cite{Carvalho2022,Allen2022}.
    \item \textbf{Vesting}: releasing tokens gradually to avoid sudden supply shocks from early stakeholders~\cite{Yoo2021,Li2018}.
\end{itemize}

\textit{Design implication and trade-off.} Supply-reduction mechanisms can support price stabilization but can also amplify future volatility if large unlock events occur or if staking yields attract primarily speculative participants. Accordingly, TEDM treats price management as a toolbox whose suitability depends on the token's value-capture channel and anticipated adverse scenarios.

\textit{Currynomics.} Redcurry's primary value anchor is its link to real assets and the associated trust dynamics. A complementary channel is network value: broader adoption can reinforce perceived credibility for new users. For the DAO token, key value channels are governance rights and (prospectively) claims on earnings, insofar as transaction fees may indirectly support token value over time. Regarding price management, Redcurry incorporates a redemption/buyback-style control logic tied to demand changes; in addition, users can stake Redcurry and receive DAO tokens as a reward, which reduces circulating supply and can dampen short-run sell pressure. For the DAO token, vesting is applied to mitigate early-contributor sell-offs, and the chosen voting mechanism provides an additional lock-up channel that can reduce circulating supply during governance participation. The team is considering whether additional price-management mechanisms will be necessary as the ecosystem matures and as liquidity conditions evolve.

\section{Evaluation}\label{sec:evaluation}
This section presents the evaluation of the Token Economy Design Method (TEDM) following the iterative DSR cycles described in Section~\ref{sec:methdology}. The evaluation is \emph{formative}: it assesses TEDM's interpretability, completeness, simplicity, understandability, operational feasibility, and perceived accuracy, rather than providing statistical generalization or causal proof of superior economic outcomes. Since the Currynomics ecosystem had not yet launched at the time of writing, outcome-oriented validation (e.g., long-run sustainability, robustness under market stress) is outside the scope of the empirical evidence available in this version of the study. To partially address external validity and failure-mode concerns, we complement the focal-case demonstration with (i) external expert interviews and (ii) a structured, desk-based stress discussion that links TEDM decision points to common adverse scenarios.

First, we report the use-case \textit{demonstration} in Section~\ref{subsec:use_case_demonstration}. Second, we present the case-based evaluation through interviews with Currynomics participants in Section~\ref{sec:case_study_evaluation}. Case-based evaluation is widely adopted in DSR studies~\cite{Peffers2012}; however, the maturity of the focal system constrains what can be validated empirically at the time of writing. Therefore, a further evaluation cycle is conducted through semi-structured interviews with industry experts not related to the Currynomics case (Section~\ref{expertinterviews}). Finally, Section~\ref{subsec:thematic_analysis_overview} provides an overview of the thematic-analysis coding results.

\begin{table}[h]
    \centering
    \begin{tabular}{|c|l|p{8cm}|}
        \hline
        \textbf{Code} & \textbf{Role} & \textbf{Background} \\ 
        \hline
        IN1 & Co-founder of the Redcurry Foundation & Business development, MBA \\ 
        \hline
        IN2 & Co-founder of the Redcurry Foundation & Engineer, BSc in Information Technology \\ 
        \hline
        IN3 & External Industry Expert & Electrical engineer, researcher in token engineering. \\ 
        \hline
        IN4 & External Industry Expert & Web3 project consultant and advisor. \\ 
        \hline
        IN5 & External Industry Expert & MSc in Computer Science, software developer, and Web3 infrastructure founder. \\ 
        \hline
    \end{tabular}
    \caption{Interviewee roles and backgrounds.}
    \label{tab:interviewees_backgrounds}
\end{table}

Experts were purposively sampled based on the criterion of having at least three years of substantial experience in token-economy design, either through industry practice or academic work. The mix of participants supports triangulation between case-specific constraints and broader practitioner and research perspectives.

\subsection{Use Case Demonstration}
\label{subsec:use_case_demonstration}
The demonstration illustrates how TEDM supports moving from environmental problem statements to structured design propositions across incentives, governance, and tokenomics.

\textbf{Incentives (Section~\ref{sec:TEDMincentives}).} TEDM supported the elicitation of (i) stakeholder categories, (ii) system functions, and (iii) desirable behaviors to be incentivized. Given these outputs, the Currynomics team concluded that incentives linked to Redcurry should primarily be monetary (reflecting a financial value proposition), while participation in governance around the DAO token should rely more on non-monetary mechanisms (e.g., governance participation and reputational signals), with monetary appreciation as a secondary incentive.

\textbf{Governance (Section~\ref{sec:TEDMgovernance}).} TEDM supported mapping governance areas to stakeholders and clarifying which decisions remain centralized due to legal/operational constraints. Based on decentralization targets and the desired voting properties, the method guided the selection of a time-weighted voting family and, given simplicity constraints, the adoption of conviction voting as an initial mechanism. The team further concluded that support mechanisms should be introduced only as governance complexity increases.

\textbf{Tokenomics (Section~\ref{sec:TEDMtokenomics}).} TEDM supported structuring token supply policy, release timing, and distribution mechanisms. For Redcurry, the method confirmed a demand-driven, collateral-constrained issuance consistent with its asset-backed design. For the DAO token, the method clarified trade-offs of capped supply (e.g., reinforcing holding incentives) and motivated the use of vesting and token-locking to mitigate early sell pressure and reduce short-run volatility. The method also made explicit how value-capture channels (asset-linked, governance rights, prospective earnings) inform price-management choices and failure-mode risks.

\subsection{Case Study-Based Evaluation} 
\label{sec:case_study_evaluation}
As described in Section~\ref{sec:methdology}, TEDM was evaluated using five criteria: completeness, simplicity, understandability, operational feasibility, and perceived accuracy. Due to space constraints, Figure~\ref{fig:tokeneconomydesignmodel} reports the refined version of TEDM after incorporating feedback.

\paragraph{\textbf{Completeness}} From the perspective of the Currynomics case, IN2 perceived TEDM as covering the major governance, incentive, and tokenomics choices relevant at the current stage, and highlighted that the next practical step would be quantitative refinement (e.g., simulations) for selected mechanisms. IN1 suggested that token distribution guidance (pre- and post-launch) and best practices around initial price formation could be expanded in future extensions of the method.

\paragraph{\textbf{Simplicity}} Both IN1 and IN2 considered the method adequately simple given the inherent complexity of token-economy design, emphasizing that the visual structure helps reduce cognitive overload without omitting essential topics

\paragraph{\textbf{Understandability}} Both participants noted that the method benefits from explicit reading guidance and ordering across the three pillars and their sub-steps. This feedback motivated the inclusion of progressive step indices and clearer navigation cues in Figure~\ref{fig:tokeneconomydesignmodel}. They also emphasized that governance concepts introduce social-science terminology that may be unfamiliar to technically oriented practitioners, reinforcing the need for concise definitions and rationale statements per step.

\paragraph{\textbf{Operational Feasibility}}
Participants perceived TEDM as practically useful for early-stage structuring and for ruling out unsuitable mechanisms under case constraints. They noted that guidance on selecting a DAO technology stack would increase operational value, while acknowledging that such implementation-level choices are beyond the current scope. A key observation was that TEDM helps focus effort by clarifying what does \emph{not} fit the case constraints, thereby narrowing the design space.

\paragraph{\textbf{Perceived Accuracy}}
IN1 cautioned that potential downsides of uncapped supply should be more prominently balanced against benefits, pointing to dilution dynamics and cascade sell-off risk. This concern aligns with prior discussions on valuation uncertainty and instability risks under unrestricted contributions~\cite{Kaal2022}. The feedback motivates more explicit boundary-condition statements in TEDM (e.g., when uncapped/demand-driven issuance is appropriate and what safeguards are required).

\subsection{Expert Interviews}\label{expertinterviews}
This subsection reports feedback from external experts after a presentation of TEDM and the case instantiation (Figure~\ref{fig:tokeneconomydesignmodel}).

\paragraph{\textbf{Completeness}}
IN3 emphasized that token-economy design is tightly coupled with legal and regulatory constraints and suggested adding an optional legal-awareness module, while acknowledging that a full legal framework is beyond the scope of this version. IN3 also stressed the need to consider incentive design through a game-theoretic lens, highlighting how poorly aligned payoff structures can trigger adverse dynamics (e.g., rapid sell-offs). IN4 was comfortable with TEDM's current scope but suggested that future work could provide modular extensions for different DAO archetypes. IN5 proposed enriching governance with staged decision-making (e.g., proposal intake, deliberation, voting, execution) and tool support by stage, consistent with views that governance evolves over time~\cite{Laatikainen2023}.

\paragraph{\textbf{Simplicity and feasibility}}
IN3 and IN5 considered the scope appropriate and did not identify clearly redundant components. IN5 appreciated the separation between core voting mechanisms and support mechanisms as a pragmatic ``add-on'' approach for later ecosystem maturity. IN4 noted that the method is practical for newcomers but requires additional reading to master details, which is expected for this domain.

\paragraph{\textbf{Perceived accuracy and boundary conditions}}
Experts emphasized that full decentralization is rare in practice and can reduce flexibility, consistent with empirical challenges observed in DAO governance~\cite{Fritsch2022}. They also argued for hybrid on-chain/off-chain designs as many effective decision-making tools remain off-chain and less costly, supporting the hybrid governance positioning in TEDM~\cite{Mosley2022}. Regarding Sybil resistance, IN3 cautioned that identity-based mechanisms can introduce new attack vectors (e.g., vote buying), suggesting that the method should present such mechanisms together with trade-offs and limitations rather than as universal solutions.

\paragraph{\textbf{Adversarial stress discussion (diagnostic use)}}
Although outcome-oriented testing is outside the available empirical scope, expert feedback and the literature motivate using TEDM diagnostically against common failure modes. We therefore relate key method decision points to adverse scenarios as follows: (i) \emph{governance capture/plutocracy} is addressed by decentralization targets (Step~3) and voting-property selection (Step~5), complemented by support mechanisms (Step~7); (ii) \emph{liquidity shocks and sell-off cascades} motivate conservative release/distribution policies (Tokenomics Steps~1--3) and price-management tool choices (Step~5); (iii) \emph{speculative dominance and incentive backfire} motivate explicitly linking desirable behaviors (Incentives Step~3) to incentive types and safeguards (Steps~4--5), and clarifying boundary conditions under which monetary incentives should be used sparingly. This discussion does not constitute empirical proof of robustness; rather, it clarifies how TEDM can be used to surface risks and design trade-offs early.

\subsection{Thematic Analysis Overview}
\label{subsec:thematic_analysis_overview}
Table~\ref{tab:model_evaluation_metrics} summarizes the thematic-analysis coding results across the five interview transcripts with regard to the evaluation criteria. Counts reflect the number of coded comment segments (codes) associated with each criterion and stance. These descriptive counts support transparency about what themes were most frequently raised; they are not intended as statistical evidence.

\begin{table}[h!]
\centering
\begin{tabular}{|l|c|c|c|}
\hline
\textbf{Evaluation Criterion} & \textbf{Supporting} & \textbf{Neutral} & \textbf{Opposing} \\ \hline
Completeness              & 3                            & 4                         & 2                           \\ \hline
Simplicity                & 7                            & 2                         & -                           \\ \hline
Understandability         & -                            & 2                         & 3                           \\ \hline
Operational feasibility   & 6                            & 8                         & -                           \\ \hline
Perceived Accuracy                  & 4                            & 11                        & 3                           \\ \hline
\end{tabular}
\caption{Summary of coded interview comments by evaluation criterion}
\label{tab:model_evaluation_metrics}
\end{table}

Overall, TEDM received a positive formative assessment. Simplicity and operational feasibility were strongly supported, reinforcing practical usefulness for early-stage structuring. Completeness and perceived accuracy received mixed feedback, with recurring suggestions to (i) make legal/regulatory boundary conditions more explicit and (ii) strengthen the link between qualitative design propositions and subsequent quantitative validation. Understandability emerged as the main area of concern, motivating the refinement of navigation cues and step ordering in Figure~\ref{fig:tokeneconomydesignmodel}.

\subsection{Comparative Analysis of Curve Finance and Uniswap Token Economies}\label{subsec:comparison}
We apply the TEDM classification introduced in Sections~3--5 as an analytic coding scheme to compare two operating token economies: Curve Finance\footnote{https://resources.curve.fi/} and Uniswap\footnote{https://app.uniswap.org/}. Both are well-established decentralized exchanges (DEXs) based on automated market makers (AMMs)~\cite{ottina2023automated}. This comparative instantiation complements the prior evaluation steps by illustrating how TEDM categories can be applied beyond the focal Currynomics case. While both cases belong to the same application class, they differ substantially in incentive and governance design, thereby providing a non-trivial test of the portability of the TEDM classification within a shared domain.

Both protocols facilitate permissionless token swaps through liquidity pools rather than order books~\cite{ottina2023automated}. Uniswap employs a constant-product AMM design (commonly formalized via the invariant \(x\cdot y = k\)), enabling swaps across a broad range of ERC-20 token pairs. Governance is handled via a DAO, where UNI token holders vote on upgrades and protocol parameters. Curve Finance specializes in stable-value asset swaps, optimizing for low slippage and reduced impermanent loss through a bonding-curve design tailored for correlated assets (e.g., stablecoins and wrapped assets)~\cite{ottina2023automated}. Curve governance is also DAO-based, but relies heavily on token locking (veCRV) that links governance influence to long-term commitment and reward boosting.

In the following, we summarize the main design features of the two token economies using the TEDM pillars: incentives, governance, and tokenomics.

\subsection{Incentives}
\begin{table}[h]
\centering
\renewcommand{\arraystretch}{1.2}
\setlength{\tabcolsep}{4pt}
\resizebox{16cm}{!}{
\begin{tabular}{|p{2.5cm}|p{6.5cm}|p{6.5cm}|}
\hline
\textbf{TEDM Step} & \textbf{Uniswap} & \textbf{Curve Finance} \\ \hline
\textbf{Value Proposition} & 
\begin{itemize}
    \item \textbf{Liquidity Providers (LPs)}: Earn trading fees by supplying liquidity to pools.
    \item \textbf{Traders}: swap tokens without intermediaries at automated market prices.
    \item \textbf{UNI Holders}: participate in governance (e.g., upgrades and parameter decisions).
\end{itemize} & 
\begin{itemize}
    \item \textbf{Liquidity Providers (LPs)}: earn trading fees and CRV incentives for providing liquidity.
    \item \textbf{Traders}: exchange stable-value assets with minimal slippage.
    \item \textbf{veCRV holders}: lock CRV to gain governance power and reward boosts.
    \item \textbf{External DeFi stakeholders}: compete to attract liquidity by influencing emissions toward specific pools.
\end{itemize} \\ \hline
\textbf{Desirable Behaviors} & 
liquidity provision; participation in governance. & 
Long-term commitment; active governance; directing incentives toward strategically relevant pools. \\ \hline
\textbf{Incentive Mechanisms} & 
Trading fees for LPs (monetary); governance rights for UNI holders (non-monetary). & 
Trading fees for LPs (monetary); CRV emissions to liquidity providers (monetary); veCRV-based boosted yields (monetary);  governance influence via veCRV locking (non-monetary).\\ \hline
\end{tabular}}
\caption{TEDM-based comparison of incentive structures in Uniswap and Curve Finance.}
\label{tab:incentives_comparison}
\end{table}

Applying TEDM to incentives highlights how both protocols align liquidity provision and governance participation, but with different mechanism complexity (Table~\ref{tab:incentives_comparison}). In Uniswap, the stakeholder set is relatively compact: LPs are rewarded via fee distribution, while UNI holders gain governance rights. Desirable behaviors are continuous liquidity provision and governance participation, supported primarily by fee-based monetary incentives and non-monetary governance rights.

Curve Finance introduces additional incentive channels and roles. Beyond fees, CRV emissions reward liquidity provision, while locking CRV as veCRV increases both governance power and reward boosts. In addition, external DeFi projects may seek to influence veCRV votes to direct CRV emissions toward their pools (often discussed as ``bribing'' dynamics)~\cite{lloyd2023emergent, allen2023exchange}. This creates a stronger coupling between incentives and governance: governance participation is not only non-monetary (decision rights) but also tightly linked to monetary outcomes via boosted yields and emissions allocation.

\subsection{Governance}
\begin{table}[h]
\centering
\renewcommand{\arraystretch}{1.2}
\setlength{\tabcolsep}{4pt}
\resizebox{16cm}{!}{
\begin{tabular}{|p{4cm}|p{6.5cm}|p{6.5cm}|}
\hline
\textbf{TEDM Step} & \textbf{Uniswap} & \textbf{Curve Finance} \\ \hline
\textbf{Governance Areas} & 
Protocol upgrades, fee-related decisions, and treasury management. & 
Liquidity incentives/emissions allocation, protocol upgrades, and treasury allocation. \\ \hline
\textbf{Stakeholder Roles} & 
UNI holders propose (via governance processes) and vote on changes. & 
veCRV holders vote on liquidity incentives and other governance decisions; LPs can influence outcomes indirectly via locking decisions and emissions-focused participation.  \\ \hline
\textbf{Level of Decentralization} & 
Permissionless but highly concentrated voting power (reported as highly centralized). & 
Permissionless; more distributed voting power than Uniswap, but still meaningfully concentrated. \\ \hline
\textbf{Voting Mechanism} & 
 1-Token-1-Vote based on UNI. & 
Time-weighted vote-escrow governance (veCRV: longer lock \(\Rightarrow\) higher vote weight). \\ \hline
\end{tabular}}
\caption{TEDM-based comparison of governance models in Uniswap and Curve Finance.}
\label{tab:governance_comparison}
\end{table}

Table~\ref{tab:governance_comparison} summarizes governance properties under TEDM. Uniswap relies on a standard 1-token-1-vote model, which is simple but vulnerable to plutocratic concentration. Prior work reports a Gini coefficient of 0.996 and a Nakamoto-style threshold of 11 for Uniswap governance power, indicating extreme concentration~\cite{Fritsch2022}. This supports low inclusivity and elevated capture risk for the governance process.

Curve Finance uses a vote-escrow model (veCRV), where voting power depends on both the amount of CRV locked and the lock duration. Our analysis reports a Gini coefficient of 0.8402 and a Nakamoto coefficient of 23, suggesting broader distribution of voting power relative to Uniswap, though still centralized in absolute terms. These metrics should be interpreted as distribution-based signals of governance concentration; they do not fully capture turnout effects, delegation patterns, or off-chain influence channels. The full methodology and dataset for the Curve concentration analysis are available in the associated GitHub repository\footnote{https:\/\/github.com\/SoweluAvanzo\/Curve\_Finance\_data\_analysis}.

\subsection{Tokenomics}

\begin{table}[h]
\centering
\renewcommand{\arraystretch}{1.2}
\setlength{\tabcolsep}{4pt}
\resizebox{16cm}{!}{
\begin{tabular}{|p{3cm}|p{7cm}|p{7cm}|}
\hline
\textbf{TEDM Step} & \textbf{Uniswap} & \textbf{Curve Finance} \\ \hline
\textbf{Token Supply Model} & 
Inflationary policy (e.g., 2\% annual inflation reported in the literature~\cite{uniswapUNI2020}). & 
Capped supply (reported cap \(\approx 3.03\)B CRV). \\ \hline
\textbf{Token Distribution} & 
Post-launch distribution via airdrop and (time-bounded) liquidity mining; allocations for treasury, team, and investors. & 
Post-launch emissions via liquidity mining; vested allocations for team and early stakeholders. \\ \hline
\textbf{Value-Capture Channels} & 
\begin{itemize}
    \item \textbf{UNI}: governance rights; network value.
    \item \textbf{LP token:} claims on pool assets and fee revenue; network value.
\end{itemize}& 
\begin{itemize}
    \item \textbf{CRV/veCRV:} governance rights; boosted rewards via locking; earnings-linked value via directing liquidity incentives.
    \item  \textbf{LP token:} claim for asset, claim for revenues, value of the network
\end{itemize}
 \\ \hline
\textbf{Price-Management Mechanisms} & 
No explicit price-stabilization mechanism; price dynamics largely market-driven (aside from vesting/lockups in allocations). & 
veCRV locking reduces circulating supply and links governance power to long-term commitment.  \\ \hline
\end{tabular}}
\caption{TEDM-based tokenomics comparison of Uniswap and Curve Finance.}
\label{tab:tokenomics_comparison}
\end{table}

Table~\ref{tab:tokenomics_comparison} highlights distinct tokenomics choices. Uniswap's UNI supply is reported as inflationary (e.g., 2\% annual inflation) without a hard cap, with an initial post-launch distribution through airdrops and liquidity mining and additional allocations to treasury, team, and investors~\cite{howell2023uniswap}. Curve Finance is reported as capped (approximately 3.03B CRV), with emissions primarily routed through liquidity mining and complemented by vested allocations~\cite{curve2020docs}. Unlike Uniswap's immediate airdrop to early adopters, CRV distribution emphasizes ongoing participation through emissions and vesting.

For both protocols, LP tokens capture value through claims on pooled assets and fee revenue, and the growth of network activity amplifies fee generation. UNI and CRV/veCRV provide governance rights; however, Curve further links governance to reward allocation and long-term locking, which can affect both incentive alignment and circulating supply dynamics. Under TEDM, these differences illustrate how value-capture channels and distribution policies interact with governance design and incentive mechanisms, supporting the use of the TEDM classification as a cross-case analytic lens (within the DEX AMM domain).

Future work should extend the comparative instantiation to structurally distinct token economies (e.g., governance-minimal protocols, stablecoins, or failed algorithmic designs) to further probe boundary conditions.

\section{Conclusion and Future Work}\label{conclusion}
This study proposed the Token Economy Design Method (TEDM), a stepwise, DSR-inspired method that supports early-stage token-economy design across three core pillars: incentives, governance, and tokenomics. TEDM is intended as a qualitative requirements, derivation and structuring method for practitioners, especially those new to Web3, who must navigate a large design space and make trade-offs explicit before proceeding to implementation and quantitative refinement.

The development of the method followed a Design Science Research approach in which prior academic knowledge was synthesized into design propositions and iteratively refined through application to a concrete case. TEDM was instantiated in the Currynomics ecosystem, which is developing an inflation-resistant, CRE-backed stablecoin (Redcurry), and evaluated formatively through semi-structured interviews with two case participants and three external experts using the criteria of completeness, simplicity, understandability, operational feasibility, and perceived accuracy. Feedback primarily motivated improvements to the method’s navigation cues and ordering, which were incorporated into the final version of the artifact.

To complement the focal case, we further instantiated the TEDM classification in a comparative analysis of two operational DeFi token economies (Uniswap and Curve Finance)~\cite{ottina2023automated}. While both belong to the DEX AMM domain, they exhibit markedly different incentive–governance couplings (fee-driven governance vs.\ vote-escrow locking and emissions direction). This demonstrates how TEDM can be applied as a reusable analytic lens to structure and compare token economy design choices beyond the focal case, within a shared application class.

Overall, TEDM contributes a method-level artifact that helps: (i) make stakeholder objectives and desired behaviors explicit; (ii) structure governance choices via decentralization targets, voting properties, and mechanism selection; and (iii) organize tokenomics decisions as a coherent policy space covering release/distribution and value/price-management mechanisms. The method is not positioned as a guarantee of economic success; rather, it supports disciplined reasoning about design trade-offs and boundary conditions, and prepares the ground for subsequent quantitative testing and refinement.

\paragraph{Limitations and Future Work}
This paragraph clarifies how the current evidence should be interpreted and outlines directions for future work. First, the evaluation is qualitative and formative. The interview-based assessment focuses on perceived usefulness and interpretability (e.g., completeness, simplicity, understandability, operational feasibility, and perceived accuracy), and should not be interpreted as causal evidence that applying TEDM leads to superior economic outcomes (e.g., sustainability, robustness, or incentive alignment) under real market conditions.

Second, the empirical basis is narrow. The in-depth evaluation relies on a small purposive sample and on a single co-designed case, which enables rich insight into design trade-offs but limits statistical generalization. The results are therefore intended to be analytically transferable rather than statistically generalizable. The comparative instantiation with Uniswap and Curve Finance supports portability of the TEDM classification beyond the focal case, but it remains within one application class (DEX AMMs) and thus does not establish cross-context validity on its own.

Third, the development of TEDM prioritizes relevance to design practice over strict reproducibility. The method was developed through an iterative synthesis of selected literature and refinement based on evaluation feedback, rather than through a fully formal and reproducible derivation pipeline. As a result, the current version provides structured guidance for qualitative requirements derivation and early design scoping, but it does not prescribe quantitative parameter choices (e.g., optimal issuance amounts) and does not provide a formal guarantee of completeness.

Fourth, TEDM’s applicability depends on context. It is most useful for token economies where incentives, governance, and token-supply policies are first-order design decisions and where stakeholder behaviors can be meaningfully shaped through such mechanisms. It may be less informative for governance-minimal systems or purely speculative token launches where design intent is not tied to sustained utility or institutional constraints. In addition, the present version abstracts from several technology-layer decisions (e.g., base-layer blockchain choices, middleware/DAO stack selection, and execution-layer constraints) that can materially affect feasible governance and tokenomics designs.

These points motivate several directions for future work. A natural extension is broader cross-context evaluation across structurally distinct token economies (e.g., lending protocols, stablecoins with different collateralization strategies, governance-minimal systems, and failed designs) to better formalize boundary conditions. Another direction is integrating simulation and state-space exploration approaches from token engineering to complement TEDM with quantitative refinement and stress testing, enabling outcome-oriented validation under adverse scenarios~\cite{Zargham2018StateSpaceBlockchain,RiceZargham2019TokenEngineeringIII,Zargham2020}. Further work can also strengthen reproducibility by adopting a more explicit derivation protocol (e.g., systematic mapping with transparent inclusion criteria and traceability from literature constructs to method steps). Future work will complement TEDM with on-chain behavioral metrics (e.g., token velocity and holding-time distributions)~\cite{kraner2025money}. This will help evaluate whether TEDM-informed choices align with observed usage and governance dynamics in operational token economies.
Finally, TEDM can be extended with optional modules addressing legal/regulatory awareness, base-layer and middleware choices, and DAO archetypes, including, for instance, protocol, investment, and social DAOs \cite{bonnet2024decentralized}, as well as additional tokenomics options such as dynamic-supply mechanisms (e.g., bonding-curve-based policies) when appropriate~\cite{Zargham2020}.
\section*{Acknowledgement}
\label{sec:acknowledgment}
This project is also partially funded by the Estonian ''Personal research funding: Team Grant (PRG)" project PRG1641.
\section*{Declaration of Competing Interests}
Authors have no competing interests to declare.
\bibliographystyle{elsarticle-num}
\bibliography{main}

\end{document}